# The spatial collection efficiency of photogenerated charge carriers in photovoltaic and photoelectrochemical devices

Gideon Segev[1,2], Hen Dotan[1], David S. Ellis[1], Yifat Piekner[1], Dino Klotz[1], Jeffrey W. Beeman[2], Jason K. Cooper[2], Daniel A. Grave[1], Ian D. Sharp[2,3], and Avner Rothschild[1*]

[1]Department of Materials Science and Engineering, Technion – Israel Institute of Technology, Haifa 32000, Israel

[2]Joint Center for Artificial Photosynthesis, Lawrence Berkeley National Lab, Berkeley, CA 94720, USA

[3]Walter Schottky Institut and Physik Department, Technische Universität München, Am Coulombwall 4, 85748 Garching, Germany

[*]avner@mt.technion.ac.il

The spatial collection efficiency portrays the driving forces and loss mechanisms in photovoltaic and photoelectrochemical devices. It is defined as the fraction of photogenerated charge carriers created at a specific point within the device that contribute to the photocurrent. In stratified planar structures, the spatial collection efficiency can be extracted out of photocurrent action spectra measurements empirically, with few *a priori* assumptions. Although this method was applied to photovoltaic cells made of well-understood materials, it has never been used to study unconventional materials such as metal-oxide semiconductors that are often employed in photoelectrochemical cells. This perspective shows the opportunities that this method has to offer for investigating new materials and devices with unknown properties. The relative simplicity of the method, and its applicability to *operando* performance characterization, makes it an important tool for analysis and design of new photovoltaic and photoelectrochemical materials and devices.

## Introduction

In photovoltaic (PV) and photoelectrochemical (PEC) cells, volume absorption of photons generates charge carriers with excess free energy, whose net flux gives rise to electric current, commonly termed the photocurrent. The spatial collection efficiency (SCE) is defined as the fraction of photogenerated charge carriers at a specific position within the cell that contribute to the photocurrent that flows out of the cell. Since the photocurrent can be used to produce electrical power or to drive an electrochemical reaction, empirical extraction of the SCE may shed light on processes that govern the energy conversion efficiency and transduction mechanisms that are important for a wide range of applications.

To date, the SCE has been used mostly as a phenomenological concept to model thin film PV cells,[1–6] photodiodes,[7] and photoelectrodes for solar water splitting.[8] In such approaches, *a priori* assumptions about the electric field distribution within the devices and drift diffusion models are commonly used to derive analytical expressions for the SCE that can be fitted to current-voltage voltammograms. While these expressions are useful for well-characterized materials and devices, applying them to new materials and devices proves difficult and frequently not possible.[9] Furthermore, the quality of the interface between different layers, which is material and process dependent, affects the electric field distribution around it. Hence, the suggested expressions for the SCE cannot be generalized for all cases; they must be tailored for different materials, structures, and processing conditions. These limitations highlight the need for an analytical method to deduce the SCE empirically, with minimal assumptions.

Electron beam induced current (EBIC) measurements are commonly used for mapping the regions in the PV cell that contribute to the current collection.[10–13] In this method, the electron beam of a scanning electron microscope (SEM) is used to generate excited charge carriers that are, in turn, collected as a measurable current for producing two-dimensional maps of the SCE. Although this method has yielded important insights into charge transport mechanisms in thin film PV cells, the need for cross section lamellas and operation in vacuum conditions make it destructive and render it difficult to evaluate devices under real operating conditions. Furthermore, EBIC measurements of solid/liquid interfaces, important for PEC cells, is practically impossible. As such, there is a pressing need for a simple, yet generalizable, method for evaluating the SCE of devices under *operando* conditions.

Extracting the SCE out of photocurrent action spectra, which are frequently measured to obtain the external quantum efficiency (EQE) of the device,[14–17] avoids most assumptions regarding driving forces and transport mechanisms, while also allowing for simple *operando* characterization of stratified planar PV and PEC devices. In PV cells with long diffusion lengths where the device thickness can be significantly larger than the wavelength of the incident photons, the charge carrier generation profile is often modeled as an exponential decay function following the Beer-Lambert law. This enables extraction of the SCE from measured photocurrent action spectra by performing an inverse Laplace transformation[14,15] or by numerical deconvolution.[16] Regularization methods were suggested to extract the SCE from EBIC measurements in which the charge carrier generation profile follows more complex functions.[18,19] These regularization methods can handle arbitrary charge carrier generation profiles, making them applicable for extracting the SCE of thin film devices, where optical interference gives rise to complex light intensity profiles that no longer follow the Beer-Lambert exponential decay behavior.[8,17] This extraction method was applied to PV devices made of well-understood materials such as silicon,[14–16] InP,[16] CuInGaSe$_2$[17] and CdS/CdTe,[14] thereby enabling validation of the extracted SCE profiles by comparing them to analytic solutions obtained by device simulations. By fitting the extracted profiles to the analytic solutions, important material properties, such as the diffusion length and surface recombination velocity, were deduced.

Although the potential strength of empirical SCE analysis lies in its ability to provide valuable information on driving forces and photocarrier properties with very few *a priori* assumptions, it has only been applied so far for conventional PV cells made of fairly well-understood materials. To this day it has never been applied to study PEC cells, which are difficult to simulate and to which methods such as EBIC cannot be applied. Moreover, to the best of our knowledge it has never been applied to study nonconventional materials with poorly understood properties. This perspective article aims to highlight the opportunities that the SCE analysis has to offer for studying elusive materials and devices. First, following prior work, the SCE is extracted from the EQE spectrum of a crystalline silicon PV cell and is compared to the analytic solution. Next, the analysis is applied to a thin film hematite (α-Fe$_2$O$_3$) photoanode for PEC water splitting. Extracting the SCE profiles under *operando* conditions provides important insights into bulk vs. surface limited photocurrents and the complex electro-optical properties of the material. The relatively simple experimental apparatus required to implement the method, together with the important insights it provides, make it an important tool for studying new materials and devices for PV and PEC cells.

## Theory

Assuming a stratified planar structure with homogenous layers, all device properties, including the SCE, change only with the distance from the surface, $z$. Figure 1 shows a cross sectional illustration of the energy band diagram of a p$^+$-n-n$^+$ PV cell made of a lossy semiconductor material operated at a voltage below the open circuit voltage. Holes that are generated in the vicinity of the p$^+$-n junction (marked ① in Figure 1) are accelerated towards the junction by the built-in field. Once injected into the p$^+$ region, holes are no longer minority carriers and are less susceptible to recombination. On the other hand, holes that are generated farther away from the p$^+$-n junction must travel a longer distance before being collected and are more prone to recombination (marked ② in Figure 1). Hence, in this example, the SCE, denoted by $\phi(z)$, has a maximum near the p$^+$-n junction and decreases with distance from it, as illustrated in Figure 1.

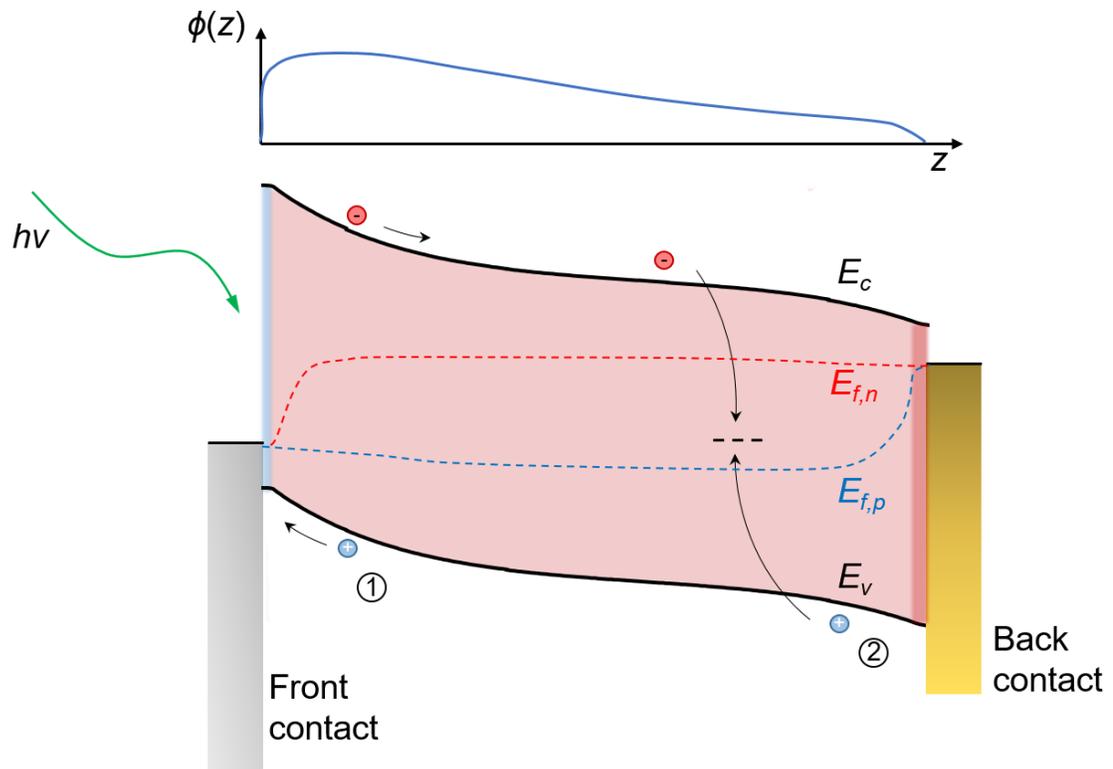

*Figure 1. Cross sectional illustration of the energy band diagram of a lossy PV cell and its SCE profile.*

*Cross sectional illustration of the energy band diagram of a lossy p$^+$-n-n$^+$ PV cell (bottom) and the corresponding qualitative SCE profile (top). Minority charge carriers that are generated near the p$^+$-n junction can be collected and injected to the front contact, yielding a high SCE in that region (marked ①). However, minority charge*

carriers generated farther away from the junction have a higher probability of recombining through bulk defects (marked ②), leading to a gradual decrease in the SCE with the distance from the junction.

The SCE is defined as the fraction of charge carriers photogenerated at point $z$ that contribute to the measurable photocurrent density, $J_{photo}$.[14–21] The relation between $J_{photo}$ and the SCE profile, $\phi(z)$, can be described as:[8]

$$J_{photo} = q \int_0^d G(z)\phi(z)dz \tag{1}$$

where $q$ is the electron charge, $d$ is the absorber layer thickness and $G(z)$ is the charge carrier generation profile. In conventional semiconductors, such as Si and GaAs, the charge carriers behave as free electrons and holes and their transport properties (e.g., mobility and lifetime) are independent of the absorbed photon energy. In this case, $G(z)$ follows the light absorption profile, $G(z) = \int \Phi_{in}(\lambda)A(\lambda, z)d\lambda$, where $\Phi_{in}(\lambda)$ is the incident photon flux at wavelength $\lambda$, and $A(\lambda, z)$ is the fraction of the incident photons with wavelength $\lambda$ that are absorbed at distance $z$ from the front surface. Since the light absorption profile $A(\lambda, z)$ can be calculated by optical modeling,[8] $\phi(z)$ can be obtained by solving equation ( 1 ). However, equation ( 1 ) has an infinite number of possible solutions and more information about the system is required in order to obtain the physical solution that characterizes the system uniquely.

One method to obtain more information on the system is to measure the photocurrent response to small perturbations to the charge carrier generation profile, for example, by modulating the intensity of the incident light at varying wavelengths on top of a constant background light bias that defines the operating point. Thus, the incident photon flux becomes $\Phi_{in} = \Phi_{white\ bias} + \Delta\Phi(\lambda)$ where $\Phi_{white\ bias}$ is the background photon flux of the light bias and $\Delta\Phi(\lambda)$ is the additional photon flux at wavelength $\lambda$. A short discussion about the background light bias requirements in EQE measurements can be found in the supporting information. $\Delta\Phi(\lambda)$ gives rise to additional photocurrent:

$$\Delta J_{photo}(\lambda) = q \int_0^d \Delta G(\lambda, z)\phi(z)dz \tag{2}$$

where $\Delta G(\lambda, z)$ is the additional charge carrier generation induced by $\Delta\Phi(\lambda)$. The EQE is defined as:

$$EQE(\lambda) = \frac{\Delta J_{photo}(\lambda)}{q\Delta\Phi(\lambda)} \tag{3}$$

Equation ( 2 ) can be rewritten in matrix form, where the unknown SCE vector, $\bar{\phi}(z_i)$, minimizes the matrix norm:

$$\epsilon = \left\| q \cdot \Delta\bar{\bar{G}}'(\lambda, z_i)\bar{\phi}(z_i) - \Delta\bar{J}_{photo}(\lambda) \right\|_2 \to 0 \tag{4}$$

Here, $\Delta\bar{J}_{photo}(\lambda)$ is a vector that is derived from the measured photocurrent action spectrum upon light intensity perturbation $\Delta\Phi(\lambda)$, $\Delta\bar{\bar{G}}'(\lambda, z_i) = \int_{z_i} \Delta G(\lambda, z_i)dz$ is a computable matrix that accounts for the changes in the charge carrier generation within the finite elements located at discrete grid positions $z_i$, and $\bar{\phi}(z_i)$ is the SCE of these elements. Hence, $\bar{\phi}(z_i)$ can be extracted from photocurrent action spectra measurements by solving equation ( 4 ). Standard regularization methods such as Tikhonov regularization[22–24] can be applied to diminish spurious effects such as measurement noise, inaccuracies inflicted by the optical modeling, and other sources of errors.[22,23] It should be noted that this type of minimization problem, often referred to as discrete ill-posed problems, has an infinite number of solutions from which only one describes the actual physics of the system.[22–24] Methods for obtaining the physical solution are described below, and additional details are provided in the supporting information.

While charge carriers behave as free electrons and holes in conventional semiconductors such as Si and GaAs, many other semiconductor materials display strong electron-phonon coupling effects that give rise to self-trapping and polaronic phenomena. Such effects, which are particularly common among emerging semiconductors envisioned for application in PEC solar cells lead to profoundly different behavior than their conventional counterparts.[25] This is often the case for transition metal-oxide semiconductors, especially those containing partially occupied *d*-orbitals in which correlation effects underlie the electronic structure and *d-d* transitions contribute to the optical absorption spectrum but not necessarily to the photocurrent.[26,27] For such materials it cannot be assumed *a priori* that every absorbed photon generates mobile charge carriers. For example, in transition metal oxides such as hematite (α-$Fe_2O_3$) and copper vanadate (γ-$Cu_3V_2O_8$), considered as potential photoelectrode candidates for PEC cells for solar water splitting, it has been reported that *d-d* transitions produce excited states which are site-localized and hence cannot be harvested efficiently as useful photocurrent.[28–30] However, other transitions such as ligand-to-metal charge transfer (LMCT) transitions give rise to mobile charge carriers that contribute more effectively to the photocurrent.[28] Thus, different types of transitions yield different probabilities of the photogenerated charge carriers to contribute to the photocurrent, such that the effective charge carrier generation function, *G*, depends not only on the amount of light absorbed but also on the type of the electronic transition induced by the absorbed photons. This leads to a wavelength-dependent charge carrier generation profile that can be written as

$$\Delta G(\lambda, z) = \xi(\lambda) A(\lambda, z) \Delta \Phi(\lambda) \qquad (5)$$

where $\xi(\lambda)$, the photogeneration yield, is the probability for the absorbed photons to generate mobile charge carriers that can contribute to the photocurrent. The different types of transitions add another level of complexity because $\xi(\lambda)$ is another unknown that must be accounted for. However, if the SCE profile is known, $\xi(\lambda)$ can be extracted by inserting equation ( 5 ) into equation ( 2 ) and solving for $\xi(\lambda)$:

$$\xi(\lambda) = \frac{\Delta J_{photo}(\lambda)}{q \Delta \Phi(\lambda) \int_0^d A(\lambda, z) \phi(z) dz} \qquad (6)$$

This leads to an empirical method to extract $\xi(\lambda)$ in order to provide additional insight into electronic structure, optoelectronic properties, and photocarrier transport, as demonstrated in the end of this article.

## Determination of $\phi(z)$ from photocurrent action spectra

### Numerical procedure

We now turn to the approach for extracting $\phi(z)$ out of the photocurrent action spectra, $\Delta \bar{J}_{photo}(\lambda)$. This is done by inserting the measured $\Delta \bar{J}_{photo}(\lambda)$ and the corresponding charge carrier generation profile, $\Delta \bar{\bar{G}}'(\lambda, z)$, obtained by optical calculations as in Dotan et al.;[8] for example, into the minimization problem presented in equation (4). Being an ill-posed problem, it has an infinite number of solutions and the unique physical solution must be carefully selected from all other possible solutions. One method to do so is to constrain the semi-norm $\|L\phi(z)\|_2$:

$$\epsilon' = \left\|q \cdot \Delta \bar{\bar{G}}'(\lambda, z_i)\bar{\phi}(z_i) - \Delta \bar{J}_{photo}(\lambda)\right\|_2 + \kappa \left\|L\bar{\phi}(z_i)\right\|_2 \to 0 \qquad (7)$$

where $L$ is either a derivative operator of any order or the identity matrix and $\kappa$ is the regularization parameter that determines the extent to which $\|L\phi(z)\|_2$ is constrained.[23] For example, when $L$ is the identity matrix, high values of $\kappa$ produce solutions in which the magnitude of the solution is constrained, and if $L$ is the first or second derivative operator, high values of $\kappa$ constrain the slope or the curvature of the solution, respectively. It should be noted that in the latter case, $L$ is a discrete approximation of the derivative operator and it does not hold information on the spatial grid. As a result, the degree in which the actual slopes and curvatures are constrained depends also on the size of the elements in the spatial grid. A short discussion on the effect that grid discretization has on the solution can be found in the supporting information.

The solution process starts with computation of a series of solutions for different values of $\kappa$. The next step is to screen out the physical solution. In the results described below, the solutions for $\bar{\phi}(z_i)$ were screened based on the basic notion that the physical solution must be confined between 0 and 1, and that it should reproduce the measured EQE spectra when inserted into equation (2). Since sharp changes in the gradient of the SCE may result in minor overshoots and undershoots in the extracted SCE profiles,[16] the acceptable lower and upper bounds for the SCE were slightly extended, with reasonable values being in the range $-0.02 \leq \phi(z) \leq 1.02$. Last, all the solutions that comply with the screening criteria were averaged at every value of $z$ and the standard deviation at every point was calculated. It is noted that other screening algorithms were suggested in the literature[22,23,31,32] and examples for some of them are discussed in the supporting information.

### Example: Crystalline silicon PV cell

The $\phi(z)$ extraction method was verified empirically by comparing photocurrent action spectrum measurements of a $p^+$-n-$n^+$ crystalline silicon PV cell fabricated in our lab to an analytic solution, as suggested in Sinkkonen et al.[14] and Tuominen et al.[15] Details of device fabrication can be found in the methods section. The details of the optical simulations and a comparison between the simulated and measured optical properties can be found in the supporting information. Figure 2(a) shows the EQE spectrum measured at short circuit. The photocurrent action spectrum, $\Delta J_{photo}(\lambda)$, was obtained from the EQE spectrum according to equation (3) and was used to extract the $\phi(z)$ profile. Figure 2(b) shows the $\phi(z)$ profile obtained by averaging all the solutions that satisfy $-0.02 \leq \phi(z) \leq 1.02$ for all values of $z$ and that reproduce the measured EQE spectrum with a relative error below 15% at every wavelength. The standard deviation between all the solutions that comply with these requirements is on the order of the thickness of the line. The markers in Figure 2(b) indicate the center of each element of the chosen grid. The shape of the $\phi(z)$ profiles shown in Figure 2(b) match very well the expectations for hyperbolic functions, as derived analytically by Sinkkonen et al.[14] and Tuominen et al.[15] and Green.[21] Near the p$^+$-n junction (next to the surface at $z = 0$), minority charge carriers are quickly swept by the electric field and are injected into a region where they become majority carriers. Hence, $\phi(z)$ is close to 1 at the junction. On the other hand, charge carriers generated in the quasi-neutral region, farther away of the junction, are transported by diffusion. As a result, the probability that these carriers recombine increases with increasing $z$, leading to a gradual decrease in $\phi(z)$ with the distance from the junction. Finally, charge carriers that are generated near the surfaces are more susceptible to surface recombination and must traverse the n$^+$ or p$^+$ regions, which have a high concentration of impurities. As a result, the SCE drops sharply near the front and back surfaces.

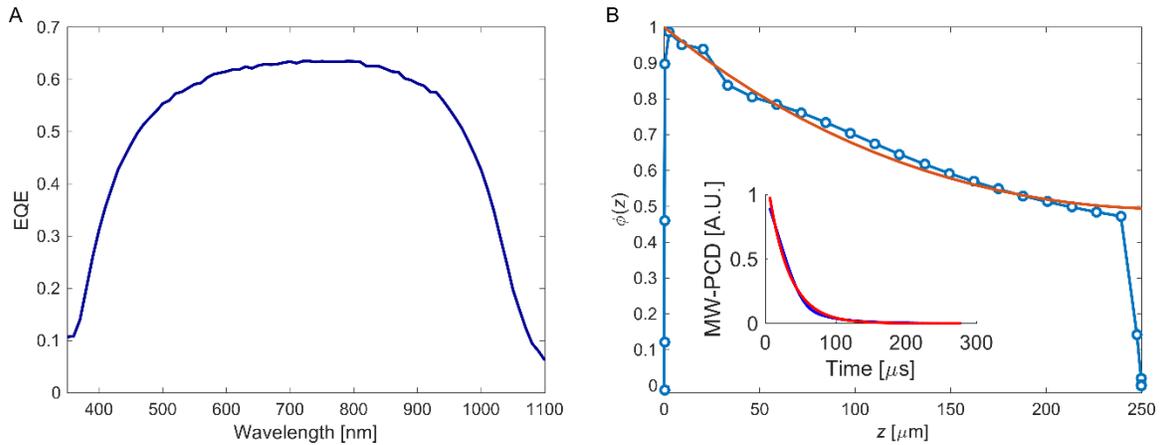

Figure 2. SCE extraction for a silicon PV cell.

(a) EQE spectrum measured for a crystalline silicon PV cell. (b) The average $\phi(z)$ profile extracted from the measured photocurrent action spectrum (blue) and the analytic solution (red) for a hole lifetime of 30 $\mu$s and surface recombination velocity of 50 cm/s. The markers indicate the center of the grid elements used in the extraction process. The surface is located at $z = 0$, from which light is incident onto the sample, i.e. near the p$^+$-n junction. The inset shows a microwave detected photoconductance decay measurement (MW-PCD, blue) for a similar wafer and its exponential fit (red).

Also shown in Figure 2(b) is the SCE profile derived from the analytic solution presented in Sinkkonen et al.[14] and Tuominen et al.[15] Assuming the space charge region at the junction is much thinner than the thickness of the wafer, the analytic solution depends only on the bulk minority carrier lifetime, $\tau_p$, the hole mobility, $\mu_p$, and the surface recombination velocity, $S_p$. The lifetime was measured by microwave detected photoconductance decay (MW-PCD), as shown in the inset of Figure 2(b). Using the extracted lifetime of 30 μs, and hole mobility of 500 cm$^2$/Vs,[33] an excellent fit is obtained with a surface recombination velocity of 50 cm/s, which is a reasonable value for a device with a back surface field (the n$^+$-n junction).[34,35] More details on the MW-PCD and the analytic solution for the SCE profile can be found in the supporting information.

Although the screening criteria used in the solution selection process are broad and generic, the standard deviation between the selected solutions is very small and the averaged SCE profile is in good agreement with the analytic solution. This indicates that the numerical solution favors converging to the physical solution, provided that it is constrained appropriately. Deviations between the analytic solution and the extracted SCE profile are noticeable near the n-n$^+$ and p$^+$-n junctions. Since the analytic solution assumes a uniform quasi-neutral region, it does not account for the sharp drops in the SCE in the highly doped regions. The deviation near the $p^+$-n junction, at $z \approx 20$ μm, and near the n-$n^+$ junction, at $z \approx 245$ μm, is a result of the sharp change in slope near this region and may be corrected with a different choice of grid.

## From well-known materials to poorly-understood ones

In the previous section, the SCE extraction method was applied to a simple device made of a well-known material (silicon) to demonstrate the concept and verify the extraction method following previous work on this topic.[14,15,17,18] However, as discussed above, the greatest potential of this method lies in its ability to probe the spatial-dependent driving forces and photocarrier properties of unconventional devices and materials with unknown electro-optical properties. To demonstrate the potential of this method to study complex materials and devices, we selected a hematite (α−$Fe_2O_3$) photoanode for water photo-oxidation as a case study.[36] Specifically, we studied a 26 nm thick heteroepitaxial 1% Ti-doped α−$Fe_2O_3$ film deposited by pulsed laser deposition on a platinum coated (0001) sapphire substrate that serves as an ideal model system as described elsewhere.[37] More details on the sample characteristics and deposition method can be found in the methods section. The optical properties of the sample were measured by spectroscopic ellipsometry and were subsequently used to calculate the reflection spectrum, $R(\lambda)$, and light absorption profile, $A(\lambda,z)$, via the transfer matrix method algorithm, as described in Burkhard et al.[38] The calculated reflection and absorption spectra, as well as their comparison to UV-VIS spectrophotometry measurements, are shown in the supporting information. Next, the PEC performance of the sample was studied by voltammetry measurements under solar-simulated illumination and EQE measurements at several bias potentials above the photocurrent onset potential. The measurements were carried out in alkaline aqueous solution (1M NaOH in deionized water) with no sacrificial reagents. Figure 3(a) shows the current density vs. applied potential (*J-E*) voltammogram measured under solar simulated illumination (black curve). Figure 3(b) shows the EQE spectra measured at the potentials marked in Figure 3(a). The markers in Figure 3(a) indicate the photocurrent obtained by integrating the EQE spectra over the solar simulator spectrum.

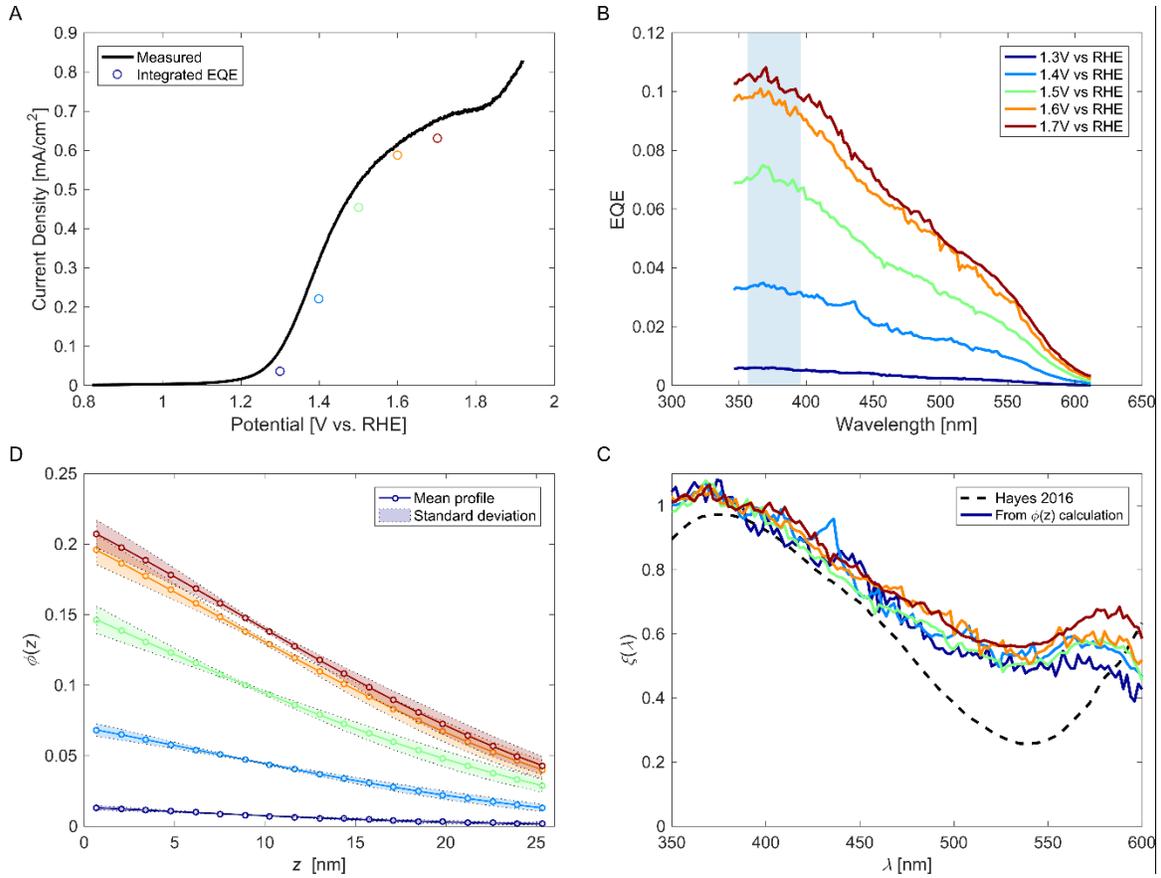

*Figure 3. SCE extraction for a thin film hematite photoanode.*

*(a) Current density vs. applied potential voltamogram measured under solar-simulated illumination (black curve), along with discrete values obtained by integrating the EQE spectra measured at the respective potentials over the spectrum of the solar simulator. (b) EQE spectra measured at the potentials marked in (a). The shaded area marks the spectral window from which the SCE profiles were extracted. (c) The SCE profiles extracted from the photocurrent action spectra in the spectral window marked by the shaded region in (b). The markers indicate the center of every element in the spatial grid. (d) The photogeneration yield spectra extracted from the SCE profiles in (c) and equation ( 6 ). Also shown is the spectrum calculated from spectroscopic results reported Hayes et al.* [28] *(black dashed curve). The color code in panels (b), (c) and (d) represents the applied potential, as denoted by the respective dots in panel (a) (see also legend in panel (b)).*

Initially, we attempted to extract SCE profiles over the full wavelength range of the photocurrent action spectra, as described above for the case of the Si PV cell. However, these efforts did not yield any solutions that comply with our selection criteria. The reason for this failure lies in the unusual electro-optical properties of hematite. As discussed in the theory section, hematite is a transition metal-oxide semiconductor whose electro-optical properties are more complicated than conventional semiconductors such as Si and GaAs. Unlike the free electrons and holes in conventional semiconductors, the charge carriers in hematite display strong electron–phonon coupling effects that lead to localization and polaronic phenomena.  Furthermore, the open-shell *d*-orbitals allow for Fe localized excitations that are ineffective at generating mobile charge carriers.[28,29] Indeed, recent studies report wavelength-dependent charge carrier dynamics and transport properties in hematite,[28,39,40] suggesting that the charge carrier generation profile depends not only on the absorption profile, $A(\lambda,z)$, but also on the photogeneration yield, $\xi(\lambda)$, that accounts for the probability that absorbed photons of wavelength $\lambda$ give rise to mobile charge carriers (see equation (5)). Therefore, extracting the SCE profile out of the photocurrent action spectra requires prior information on $\xi(\lambda)$. To overcome this barrier, the SCE can be extracted from a narrow spectral window in which the photogeneration yield can be safely assumed to be constant, as detailed below. Subsequently, $\xi(\lambda)$ can be extracted by analyzing the entire photocurrent action spectrum using the obtained $\phi(z)$ profile, as demonstrated in the following. This procedure requires a balance between using a narrow spectral window with nearly constant photogeneration yield and a wide spectral window that covers different types of transitions that give rise to both mobile charge carriers and immobile charge excitations. The implications of this tradeoff are discussed in the supporting information.

As discussed in Hayes et al.[28], the photogeneration yield of hematite changes considerably across the spectrum due to excitations that generate mobile charge carriers (ligand to metal charge transfer, or LMCT bands) and excitations that do not (*d-d* transitions), depending on the wavelength. Considering the LMCT bands reported by Hayes et al.[28], the optimal spectral window for extracting the $\phi(z)$ profile is between 356 and 396 nm, where LMCT transitions accounts for more than 93% of the total optical absorption. The SCE profiles were extracted from photocurrent action spectra within this spectral window, as indicated by the shaded region of Figure 3(b), measured at different applied potentials. The numerical procedures and selection criteria described in the previous section were applied using a grid consisting of 19 equally spaced elements. A unity photogeneration yield ($\xi(\lambda) = 1$) was assumed over the 356 – 396 nm spectral range. Figure 3(c) shows the averaged $\phi(z)$ profiles and the standard deviation at different potentials. The markers indicate the center of every element in the spatial grid. As expected for materials such as hematite that exhibit minority carrier limited transport, the SCE drops considerably with the distance from the surface. The SCE profiles follow a fairly linear shape and reach non-negligible values near the back contact, implying that the photoanode is fully depleted[41] or that surface recombination is the most dominant loss mechanism.[19] Since the reported values for the diffusion length in hematite vary from 2-4 nm[42] to 20-30 nm[43,44] and supporting measurements such as Mott Schottky analysis are problematic for such thin films,[45] it is difficult to distinguish one mechanism from the other. A study of the SCE profiles as a function of the device thickness may give more information on the collection length, the nature of the back contact, and the role of recombination on the front and back surfaces.

The SCE at the front surface, $\phi(0)$, is the probability for holes that are generated at the surface to contribute to the photocurrent. As can be seen in Figures 3(a,c), $\phi(0)$ increases with potential and then saturates at higher anodic potentials, where the photocurrent begins to plateau. On the other hand, the shapes of the SCE profiles are nearly independent of the applied potential. This indicates that the increase in potential primarily serves to increase the charge transfer efficiency, as discussed in Klotz et al.[46] rather than drive more holes to the surface, as suggested by many researchers based on the Gärtner model.[47] The values of $\phi(0)$, as shown in Figures 3(c), are lower than the charge transfer efficiencies, $\eta_t$, obtained by time and frequency domain techniques for heteroepitaxial hematite photoanodes.[48] This discrepancy stems from differences between the definitions of $\phi(0)$ and $\eta_t$. The SCE analysis gives information on the fate of photogenerated charge carriers that were born at distance *z* from the surface. Thus, $\phi(0)$ accounts only for charge carriers that were born close to the surface. On the other hand, $\eta_t$ gives information of the fate of photogenerated charge carriers that arrive at the surface, mostly from within the bulk of the photoanode. The observation of low $\phi(0)$ values (Figure 3(c)) suggests that quite significant fraction of the charge carriers that were created near the surface did not contribute to the photocurrent because they traveled in the backward direction, as discussed elsewhere.[8] This indicates that the photoanode displays poor asymmetry for charge transport, the salient driving force that gives rise to charge separation in solar cell devices,[49] possibly due to overlapping depletion regions at the front side and backside of the film.[50]

Assuming the SCE profiles at different potentials are independent of the excitation wavelength, the SCE profiles extracted in the 356 to 396 nm spectral window can be used to determine the photogeneration yield spectrum, $\xi(\lambda)$. This is accomplished by inserting the extracted SCE profiles into equation ( 6 ). Figure 3(d) shows the photogeneration yield spectra produced with the extracted SCE profiles, overlaid with expected spectrum based on the results reported by Hayes et al.[28] The qualitative agreement between the $\xi(\lambda)$ spectra is remarkable considering the simplifications in the respective analyses and the expected variations between the two photoanodes. The higher $\xi(\lambda)$ at wavelengths near 530 nm may be a result of titanium substitutions that have been implicated in reducing losses associated with *d-d* excitations, as discussed by Kim et al.[51] It is noteworthy that the extracted $\xi(\lambda)$ spectra are independent of the applied potential for most wavelengths, except for a narrow region between approximately 550 and 600 nm, where $\xi(\lambda)$ increases with the potential. This observation is in agreement with previous reports on potential-dependent absorption at 580 nm, as discussed elsewhere.[52]

It should be noted that changing the details of the numerical procedure (for example, the grid discretization or the constrained quantity) can result in fluctuations in the extracted SCE profiles. Yet, the produced solutions can be screened by their ability to reproduce the photogeneration yield as shown in Figure 3(d) and according to their magnitude and oscillatory behavior. A discussion on the effects of the numerics on the extracted profiles can be found in the supporting information.

## Challenges and opportunities

Empirical extraction of SCE profiles from photocurrent action spectra measurements under *operando* conditions can yield important information on the driving forces and photocarrier properties of semiconductor materials and photodiode devices. Although this type of analysis has already been performed on relatively simple devices made from conventional semiconductor materials whose properties are well understood, it has never been used as a tool to characterize elusive materials and complex devices. For example, applying the method to metal-oxide semiconductors that are being pursued as potential candidates for PEC solar cells can provide important insight into their transduction mechanisms and serve as tool to study photocarrier properties under *operando* conditions. The spatial information embedded in the SCE profiles can be used to advance understanding regarding the role of overlayers and underlayers, extract the charge carrier diffusion length, and reveal the origins of critical losses in these devices. Thus, the method described here represents an important tool that can be broadly applied for rational design and optimization of materials and devices.

SCE analysis can be applied to homogenous planar stratified structures where the optics can be modeled properly and the photocurrent can be assumed to flow in one direction. Careful consideration must be exercised when studying three-dimensional complex structures such as bulk heterojunction devices, where materials are randomly blended. The ability to produce fine features in the $\phi(z)$ profile is determined by the wavelength-resolution of the photocurrent action spectra and the wealth of data they carry. For example, to obtain detailed information on thick samples, different incident angles and wavelengths should promote charge carrier generation in different regions in the sample. In cases where it is difficult to obtain detailed information from the photocurrent action spectra, as in the hematite photoanodes discussed above, uncertainties may also limit the ability to identify fine features in the $\phi(z)$ profile. In such cases, prior knowledge of the system that, for example, supply the numeric algorithm with an initial guess for the $\phi(z)$ profile or provide bounds for it, may help reduce uncertainties and yield solutions closer to the physical $\phi(z)$ profile.

An important challenge limiting the application of the extraction method is the need for accurate optical modeling. Since the spatial information is extracted from the optical modeling, it must be capable of accurately describing the charge carrier generation profiles at the desired length scales, i.e. account for interference patterns, roughness, etc. Errors in the optical modeling propagate directly to the extracted $\phi(z)$ profile. For this reason, the utilization of the method for analysis of complex systems requires accurate modeling of the optics of each and every layer in the optical stack. For example, rough substrates such as fluorine-doped tin oxide (FTO) coated glass give rise to light scattering that must be accounted for by the optical models.[53] On the other hand, multiple reflections between different components of the device require accounting for incoherent optics. Refinement of optical models over different length scales will allow even broader utilization of the method.

To extract wavelength dependent SCE profiles, as in the case of the hematite photoanode presented here, it is advisable to analyze distinct photocurrent action spectra and photogeneration profiles. This can be achieved by changing the incident angle and analyzing both front- and back-illuminated photocurrent action measurements in bifacial devices. Such analysis can tell whether the charge carrier properties are wavelength dependent and yield important insights regarding operational mechanisms and electro-optical properties.

An interesting rule of reciprocity relates the SCE profile to the excess minority carrier concentration of PV cells in the dark.[54–57] As shown in these works, $\phi(z) = u(z)/u(0)$, where $u(z)$ is the normalized minority carrier concentration at distance $z$ from the junction. Applying this rule of reciprocity to PEC requires extending it to operation under potential and light biases. However, since even elementary techniques such as Mott-Schottky analysis prove difficult in thin films,[58] broadening this relationship can provide significant opportunities for studying new materials and devices. For example, understanding how the minority carrier distribution changes with potential or pH can quantify Fermi level pinning effects at the semiconductor electrolyte interface.

## Conclusions

Spatial collection efficiency analysis can yield a wealth of information about the electro-optical properties, charge carrier transport, and driving forces in photovoltaic and photoelectrochemical devices. This Perspective article presents a method to extract the SCE out of photocurrent action spectra measurements combined with optical modeling in stratified planar structures. The analysis method was demonstrated by comparing the extracted spatial collection efficiency profile of a crystalline silicon photovoltaic cell to the respective analytic solution. The analysis was also applied to a heteroepitaxial hematite photoanode, yielding both the SCE profile and the photogeneration yield spectrum. The relatively simple apparatus and the very few *a priori* assumptions required to obtain such a wealth of information make this method a key step in progressing research and development of new materials and devices for PV and PEC solar cells.

## Experimental Procedures

### Crystalline silicon PV cell fabrication

The crystalline silicon PV cell was fabricated by implanting highly doped p-type and n-type regions on the two sides of a silicon wafer. The silicon wafer was a double side polished, n-type, and (100) oriented with a bulk n-type resistivity of 2 Ωcm, corresponding to a donor concentration of approximately $2\times10^{15}$ cm$^{-3}$. The highly doped p-type and n-type regions were formed by ion implantation of $^{11}$B and $^{31}$P to doses of $4\times10^{14}$ cm$^{-2}$ and $5\times10^{14}$ cm$^{-2}$, respectively, each at an acceleration voltage of 15 kV. 100 nm thick Au contacts were evaporated through a shadow mask on both front and back surfaces of the wafer to define the active area.

### Hematite photoanode fabrication

Deposition of the heteroepitaxial Pt(111)/Fe$_2$O$_3$(0001) films on a (0001) sapphire (Al$_2$O$_3$) substrate was performed as follows. Prior to deposition, the sapphire substrate was ultrasonically cleaned with soap, acetone, ethanol, and deionized water, followed by dipping in piranha solution (3:1 H$_2$SO$_4$ : H$_2$O$_2$ by volume) and deionized water. The sample was then loaded into the vacuum chamber of the pulsed laser deposition (PLD) system (PLD/MBE 2100, PVD Products), and pumped to a base pressure of $1\times10^{-7}$ Torr. The Pt film was deposited via RF magnetron sputtering from a 50 mm diameter target of pure (99.99%) Pt (Birmingham Metal). The deposition was performed under 5 mTorr Ar pressure, 30 W forward power, and source-to-substrate distance of 75 mm. The platinum deposition was performed at a set-point temperatures of 500°C. The deposition rate was approximately 0.5 Å/ s. Directly after the platinum deposition, the sample was subjected to a 2 h anneal at a set-point temperature of 900°C under 5 mTorr Ar within the PLD chamber. Subsequent to platinum deposition and annealing, the hematite film was deposited by PLD from a 1 cation% Ti-doped Fe$_2$O$_3$ target. The hematite film was deposited using a PLD system equipped with a KrF (248 nm) excimer laser (COMPexPro 102, Coherent, GmbH). The hematite film was deposited at a set-point temperature of 700°C with a laser fluence of approximately 1.1 J cm$^{-2}$, repetition rate of 3 Hz, source-to-substrate distance of 75 mm, and oxygen partial pressure of 10 mTorr. Additional information including materials and electrochemical characterization, can be found in Grave et al.[37]

### External quantum efficiency measurements

External quantum efficiency (EQE) measurements for the silicon PV cell were carried out using a Newport 300 W ozone-free Xe lamp, from which the optical output was passed through an Oriel Cornerstone 130 1/8m monochromator. The sample current was measured with a Gamry Reference 600 potentiostat. The monochromatic light was stepped in 10 nm wavelength intervals and chopped at a period of 1 s. A Mightex GCS-6500-15-A0510 white light emitting diode and a Mightex LGC-019-022-05-V collimator were used to produce the background light bias. $\Delta J_{photo}(\lambda)$ was calculated by reducing the current generated under background light illumination from the current generated in the presence of both monochromatic and background light illumination. The incident optical output at each wavelength was measured with a Thorlabs DET100A photodiode.

The EQE measurements of the hematite photoanode in the PEC cell were measured in similar fashion as above, but at 2 nm wavelength intervals. The light chopping period was varied based on the transient response at the different potentials.  A light bias of the approximate equivalent of 0.5 Sun was applied with a white LED (Mightex GCS-6500-15-A0510).  In order to minimize the effect of current drift due to bubbles forming at the hematite/electrolyte interface, the background and monochromatic response currents where measured sequentially for each wavelength.  Nevertheless, drift errors and optical power fluctuations are estimated to contribute to a random error of up to 5% of the total EQE.  In addition, optical alignment errors may lead to systematic errors of up to 5% of total EQE. For the potentials measured, aside from the highest and lowest, the integrated EQE with the solar spectrum agrees to within 5% of the observed photocurrent.

### Optical Characterization

The silicon transmission and reflectance measurements were taken with a Shimadzu SolidSpec-3700 UV/Vis/NIR spectrometer using an integrating sphere. The baseline for the reflectance measurement was collected with an Ocean Optics STAN-SSH-NIST NIST traceable reference mirror.

The optical parameters of the hematite film and Pt layer were extracted with a J.A. Woollam M-2000 variable angle spectroscopic ellipsometer. The reflectance of the hematite photoanode was measured with an Agilent Technologies Cary series UV Vis NIR spectrophotometer.

### Bulk lifetime measurement with microwave-detected photoconductivity decay

The silicon bulk photoexcited lifetimes were characterized using microwave photoconductivity (MWPC) in a reflection geometry with backside excitation illumination by a 1064 nm, 5-7 ns pulse width, 10 Hz laser (Minilite, Continuum) with an energy of 80 nJ/pulse and an illumination area of approximately 1 cm$^2$.  The microwave probe was generated using a mechanically tuned Gunn diode oscillator at 18 GHz (74 μeV) (Cernex CMG2838-3810-01) operated at 18 mW.  The microwave signal was detected with a CFD264080055 (Cernex) and recorded on a 500 MHz oscilloscope (DPO 4054, Tektronix).

The Silicon sample was measured in 0.1M methanol/quinhydrone solution.[59] Sample preparation included 10 min of sonication in water, acetone and isopropanol, followed by 1 min of etching in 5% HF. The sample was placed in the methanol/quinhydrone solution for 30 min prior to the measurements.

### Acknowledgment

This research was supported by the European Research Council under the European Union's Seventh Framework programme (FP/200702013) / ERC (grant agreement n. 617516). D. A. Grave acknowledges support by Marie-Sklodowska-Curie Individual Fellowship no. 659491. The results were obtained using central facilities at the Technion's Photovoltaics Laboratory, supported by the Adelis Foundation, the Nancy & Stephen Grand Technion Energy Program (GTEP) and by the Solar Fuels I-CORE program of the Planning and Budgeting Committee and the Israel Science Foundation (Grant n. 152/11). The numerical procedures and the PV cell fabrication and characterization were performed by the Joint Center for Artificial Photosynthesis, a DOE Energy Innovation Hub, supported through the Office of Science of the U.S. Department of Energy under Award Number DE-SC0004993.

# Figure legends

*Figure 1. Cross sectional illustration of the energy band diagram of a lossy PV cell and its SCE profile.*

*Cross sectional illustration of the energy band diagram of a lossy $p^+$-n-$n^+$ PV cell (bottom) and the corresponding qualitative SCE profile (top). Minority charge carriers that are generated near the $p^+$-n junction can be collected and injected to the front contact, yielding a high SCE in that region (marked ①). However, minority charge carriers generated farther away from the junction have a higher probability of recombining through bulk defects (marked ②), leading to a gradual decrease in the SCE with the distance from the junction.*

*Figure 2. SCE extraction for a silicon PV cell.*

*(a) EQE spectrum measured for a crystalline silicon PV cell. (b) The average $\phi(z)$ profile extracted from the measured photocurrent action spectrum (blue) and the analytic solution (red) for a hole lifetime of 30 $\mu s$ and surface recombination velocity of 50 cm/s. The markers indicate the center of the grid elements used in the extraction process. The surface is located at z = 0, from which light is incident onto the sample, i.e. near the $p^+$-n junction. The inset shows a microwave detected photoconductance decay measurement (MW-PCD, blue) for a similar wafer and its exponential fit (red).*

*Figure 3. SCE extraction for a thin film hematite photoanode.*

*(a) Current density vs. applied potential voltamogram measured under solar-simulated illumination (black curve), along with discrete values obtained by integrating the EQE spectra measured at the respective potentials over the spectrum of the solar simulator. (b) EQE spectra measured at the potentials marked in (a). The shaded area marks the spectral window from which the SCE profiles were extracted. (c) The SCE profiles extracted from the photocurrent action spectra in the spectral window marked by the shaded region in (b). The markers indicate the center of every element in the spatial grid. (d) The photogeneration yield spectra extracted from the SCE profiles in (c) and equation ( 6 ). Also shown is the spectrum calculated from spectroscopic results reported Hayes et al. [28] (black dashed curve). The color code in panels (b), (c) and (d) represents the applied potential, as denoted by the respective dots in panel (a) (see also legend in panel (b)).*

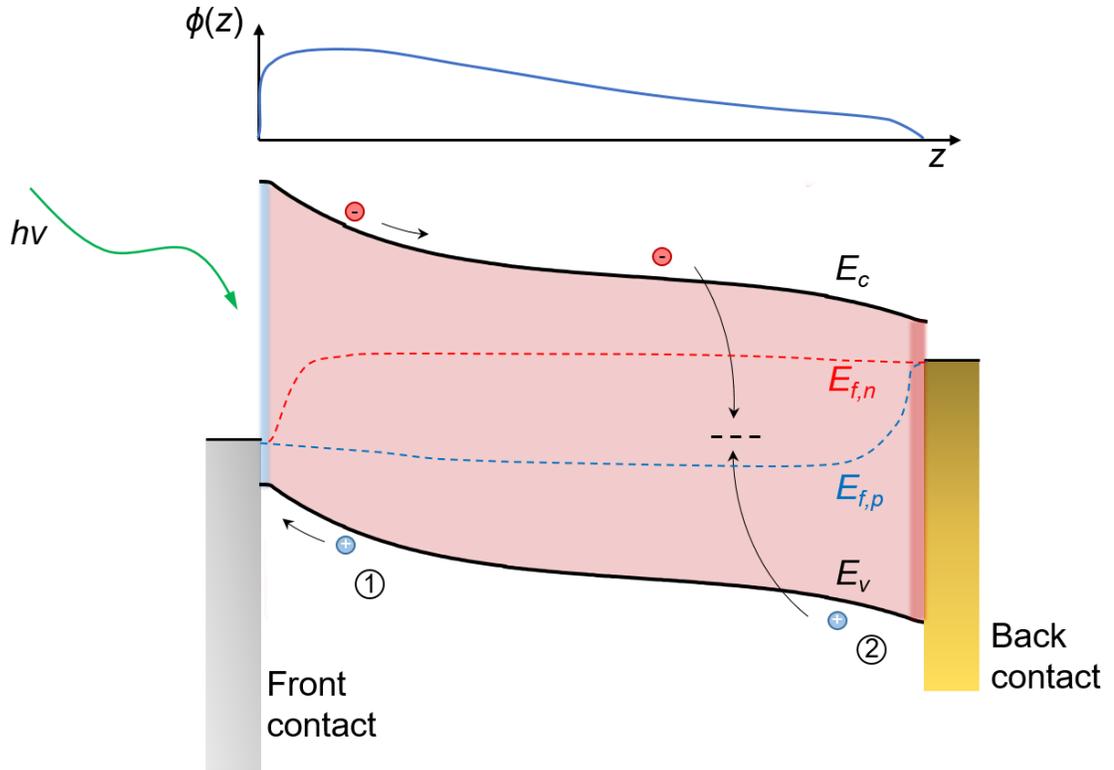

Figure 1.

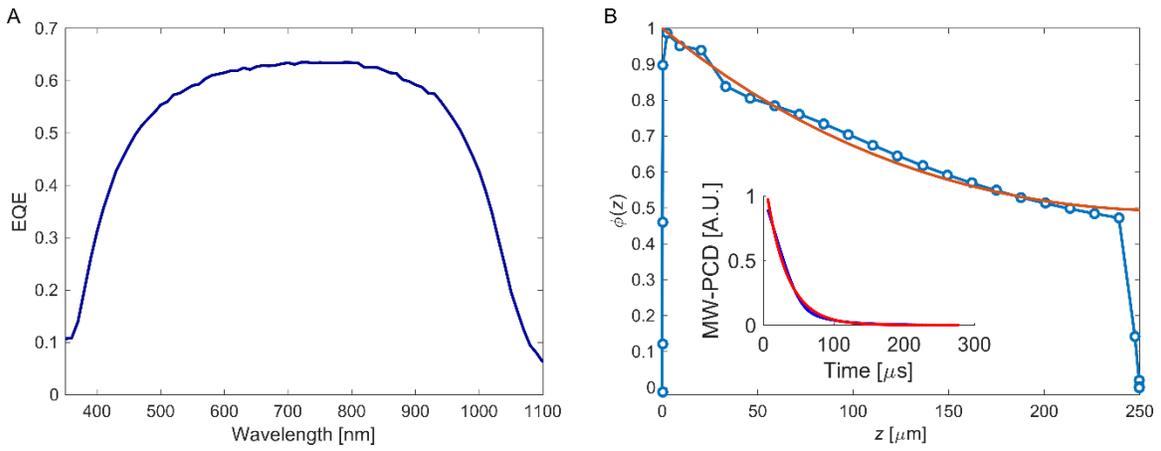

Figure 2.

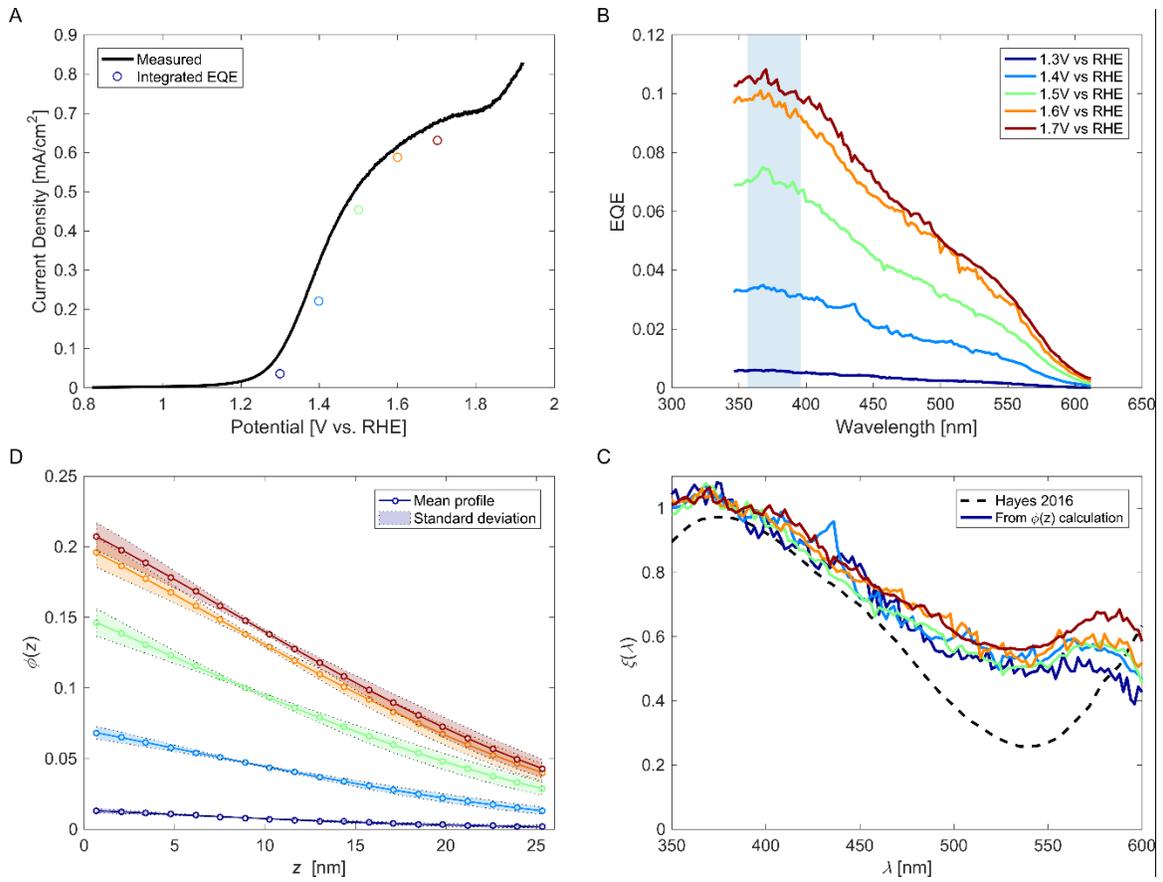

Figure 3.